\documentclass[prb,amsmath,amssymb,amsfonts,preprint]{revtex4-1}

\usepackage{graphicx}
\usepackage{natbib}
\usepackage[colorlinks=true,citecolor=red,linkcolor=blue]{hyperref}

\usepackage{verbatim}

\begin{document}
\title{\textbf{Surface plasmon-polaritons in periodic arrays of V-grooves strongly coupled to quantum emitters}}

\author{Adam Blake}
\affiliation{Department of Physics, Arizona State University, Tempe, Arizona 85287}

\author{Maxim Sukharev}
\email{maxim.sukharev@asu.edu}
\affiliation{Science and Mathematics Faculty, College of Letters and Sciences, Arizona State University, Mesa, Arizona 85212}

\date{\today}
\begin{abstract}
We investigate the optical response of a system consisting of periodic silver V-grooves interacting with quantum emitters. Two surface plasmon-polariton resonances are identified in the reflection spectrum of bare silver grooves, with the intensity of one resonance being localized near the bottom of the groove and that of the other resonance being distributed throughout the entire groove. The linear response of the hybrid silver-emitter system is thoroughly analyzed by considering the coupling between surface plasmon polaritons and emitters as the geometry of the grooves and the spatial distribution of emitters within the grooves are varied. The nonlinear response of the system is also considered by pumping the emitters with a short, high-intensity pulse. By changing the duration or the intensity of the pump, the population of emitters in the ground state at the end of the pump is varied, and it is found (upon probing with a short pulse) that an increase in the fraction of emitters in the ground state corresponds to an increase in Rabi splitting. Spatial variations in the ground state population throughout the emitter region are shown to be a result of field retardation.
\end{abstract}
\maketitle

\section{Introduction}
\label{sec:Intro}

Surface plasmon-polaritons (SPPs) are electromagnetic excitations resulting from the coupling of the incident radiation with collective oscillations of conductive electrons near the interface between metal and dielectric. Due to the specific dispersion the effective wavelength of SPPs is usually significantly shorter than that of the incident field making SPPs highly attractive for various applications \cite{Zayats2005131,ADMA:ADMA200700678}. These range from electromagnetic energy transport at the nanoscale \cite{Han:2013aa}, nano-focusing \cite{Gramotnev:2014aa}, utilization of plasmons to achieve lasing beyond the diffraction limit \cite{Noginov:2009aa}, and many others \cite{Stockman:11}. There is also a growing interest in optics of hybrid nano-materials, structures composed of plasmon-polariton sustaining systems (such as nanoparticles, one-dimensional and two-dimensional periodic arrays, for instance) and molecular aggregates \cite{Torma}. The original interest was geared towards electromagnetic energy exchange between semi-classical SPPs and molecular excitons \cite{ANIE:ANIE200903191} occurring on a femtosecond timescale \cite{Vasa2013}. Much work has been done to investigate linear optical properties of both propagating \cite{Salomon2009,1367-2630-10-6-065017,Vasa2010,Salomon2012,DeLacy:2015ab} and localized SPPs \cite{PhysRevLett.99.136802,doi:10.1021/nl4014887,doi:10.1021/ph500032d} coupled to quantum emitters in the strong coupling regime (for a comprehensive review of the current state of the field see [\onlinecite{Torma}]). Going beyond the linear regime it was recently shown that such systems present an interesting opportunity to actively control light at the nanoscale and manipulate optical properties of nano-materials \cite{jcp_chirps14,sukharev2015linear}.

Our major point of interest in this work is to scrutinize electromagnetic properties of molecular aggregates composed of simple two-level emitters strongly coupled to SPP waves supported by plasmonic waveguides. We chose a periodic one-dimensional array of V-grooves as an example of such a waveguide. V-grooves are host to various optical phenomena, many of which involve the intense, highly localized fields that result from SPP waves. Along these lines, the manipulation of channel plasmon-polaritons (CPPs) propagating along metallic V-groove channels has been extensively studied \cite{Cuesta2009, Gramotnev2004, Bozhevolnyi2010}. The propagation length of CPPs was investigated in [\onlinecite{Cuesta2009}] by filling the grooves with fluorescent microscopic beads and measuring the propagation length by microscope, and selection of CPP modes can be accomplished by adjusting the shape and size of the groove \cite{Gramotnev2004}. Additionally nanofocusing of light, which is of extreme practical interest, was demonstrated using tapered V-shaped waveguides \cite{Bozhevolnyi2010}. 

Theoretical studies have been conducted using the Green's Function Integral Equation Method \cite{EncNanotech}, yielding well-defined reflection features that are thought to result from surface plasmon polaritons, geometrical resonances, and a Wood's anomaly \cite{Sondergaard2009}. The origins of these features are identified by their response to changes in the geometry of the grooves: period, depth, groove angle, and angle of the incident fields. One point of interest is that one type of resonance displays intensity that is distributed over the entire groove whereas the intensity of another type is localized near the bottom of the groove. In this manuscript we numerically investigate how different SPPs interact with molecular aggregates in both linear and nonlinear regimes. The latter is considered using pump-probe simulations \cite{Sukharev2013}.

\section{Model}
\label{sec:Physics} 

Classical Maxwell's equations are used to describe the propagation of electromagnetic (EM) waves
\begin{subequations}
 \begin{eqnarray}
\mu_0\frac{\partial \vec{H}}{\partial t}&=&-\nabla\times\vec{E}, \label{Faraday}\\
\epsilon_0\frac{\partial \vec{E}}{\partial t}&=&\nabla\times\vec{H} - \vec{J}, \label{AmpereMaxwell}
 \end{eqnarray}
\end{subequations}
where $\mu_0$ and $\epsilon_0$ are the permeability and the permittivity of free space, respectively, $\vec{E}$ is the electric field, $\vec{H}$ is the magnetic field, and $\vec{J}$ is the current density either in the metal or that due to induced polarization of the emitters as discussed below.

\begin{figure}
\begin{center}
\includegraphics[width=0.5\textwidth]{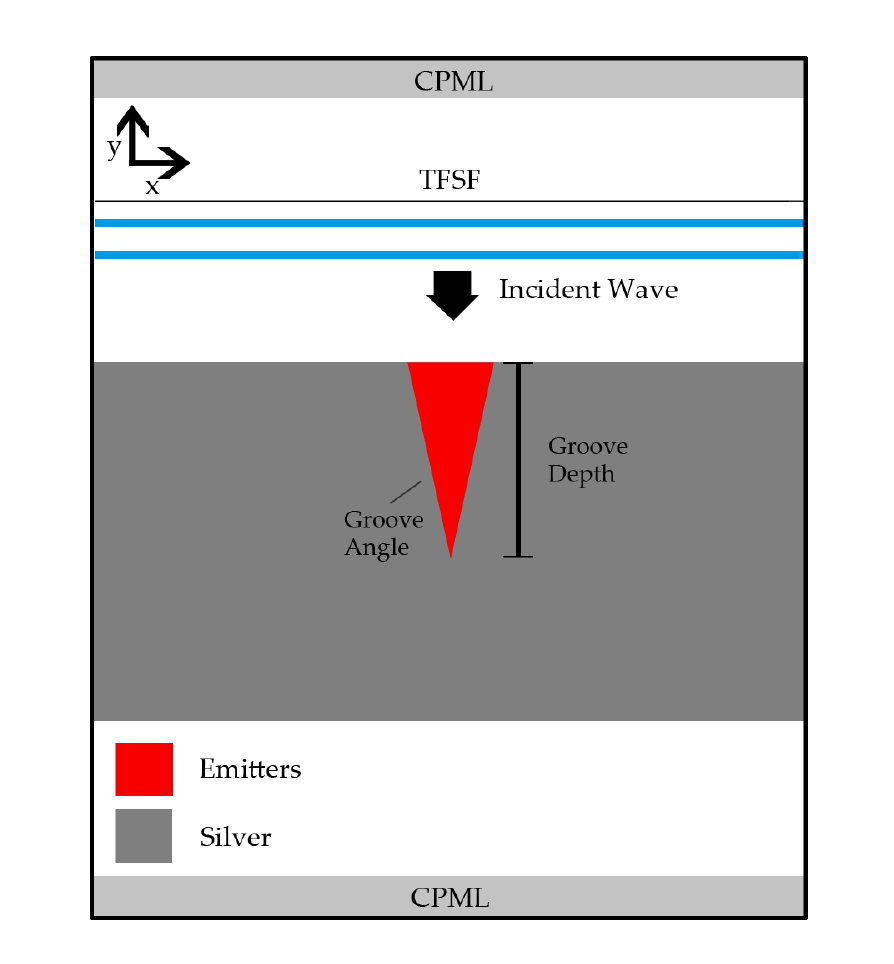}
\caption{\label{figure1} Schematic diagram of the system. The absorbing boundary conditions are implemented as \textit{Convolutional Perfectly Matched Layers} (CPML), the periodic boundary conditions are applied along $y$-axis, and the incident wave is introduced via a \textit{total-field / scattered-field} (TFSF) approach.}
\end{center}
\end{figure}

The finite-difference time-domain (FDTD) method is used to propagate Eqs. (\ref{Faraday}) and (\ref{AmpereMaxwell}) in space and time \cite{Taflove}. All of the results presented herein use a two-dimensional grid in which the fields $E_x$, $E_y$ and $H_z$ are evaluated in the $x$-$y$ plane and the structure is taken to be infinitely long in the $z$-direction, as shown in Fig. \ref{figure1}. 

The system under consideration is open in the $y$ direction and periodic in $x$. We add absorbing boundaries using \textit{convolutional perfectly matched layers} (CPML) on the top and the bottom of the grid as shown in Fig. \ref{figure1}. The left and right sides of the grid are terminated with periodic boundary conditions (PBCs) and the incident wave is introduced via a total field / scattered field (TFSF) approach \cite{Taflove}. A spatial step of $1.0$ nm was selected as results are converged for this spacing, which was demonstrated by obtaining the same data using a spatial size of $0.5$ nm. A time step of $dx/(2c)$ was chosen such that the Courant stability condition is satisfied.

The linear Drude model is used to describe the dispersion of the metal \cite{Uwe}
\begin{equation}
\epsilon(\omega)=\epsilon_r - \frac{\omega_p^2}{\omega^2-i\gamma\omega}, \label{DrudePermittivity}
\end{equation}
where $\omega_p$ is the plasma frequency, $\gamma$ is the phenomenological damping, and $\epsilon_r$ is the high-frequency limit of the dielectric function. For silver, we use the following parameters \cite{Gray2003}: $\omega_p = 11.59$ eV, $\gamma = 0.2027$ eV, and $\epsilon_r = 8.26$. In metal, the dynamics of the current density $\vec{J}$ satisfy the following equation\cite{Gray2003}:
\begin{equation}
\frac{\partial \vec{J}}{\partial t}=-\gamma\vec{J}+\epsilon_0\omega_p^2\vec{E}.
\end{equation}

In the linear regime, when the frequency response of a system to external EM excitation is independent from the incident intensity, one can use a short pulse method to obtain the spectrum of the system within a single FDTD run \cite{Sukharev2011}. Under the assumption that only the elastic scattering contributes to the spectrum, the reflection is calculated by launching a short pulse (of duration $\tau = 0.15 fs$ and whose form is $E_0 \sin^{2}(\frac{\pi t}{\tau}) \cos(\omega t)$) and spatially integrating the $y$-component of the Poynting vector (formed by the cross product of the Fourier transformed E- and H-fields; specifically, $E_x H_z$) along a line of constant $y$-value on the input side. This calculation is performed in the scattered-field region, thereby measuring only the reflected fields. 

The time dynamics of the interaction between the EM field and the molecular aggregate is described by the Liouville-von Neumann equation
\begin{equation}
i\hbar\frac{d\hat{\rho}}{dt}=[\hat{H},\hat{\rho}]-i\hbar\hat{\Gamma}\hat{\rho},
\end{equation}
where $\hat{\rho}$ is the single-emitter density matrix, $\hat{\Gamma}$ describes relaxation processes, and $\hat{H}$ is the Hamiltonian describing an emitter with a dipole moment operator $\hat{\vec{\mu}}$ interacting with an electric field
\begin{equation}
\hat{H}=\hat{H}_0-\hat{\vec{\mu}}\cdot\vec{E}(t).
\end{equation}
The expectation value of the dipole moment operator is obtained by evaluating $\text{Tr}(\hat{\rho}\hat{\vec{\mu}})$. The updated expectation value of the dipole moment is used to calculate the macroscopic polarization, $\vec{P}$, and then the polarization current, $\vec{J_p}$ , which is subsequently inserted into Eq. (\ref{AmpereMaxwell})
\begin{subequations}
 \begin{eqnarray}
\vec{P}&=&n_a\left<\vec{\mu}\right>, \\
\vec{J_p}&=&\frac{\partial \vec{P}}{\partial t}, 
 \end{eqnarray}
\end{subequations}
where $n_a$ is the volume density of emitters. 

In order to account for all possible local field polarizations we consider quantum emitters with three energy levels: an $s$-type ground state and two degenerate excited $p$-type states. In the basis of angular momentum, wave functions are chosen \cite{Sukharev2011}: $\left |1\right>=\left |s\right>$, $\left |2\right>=(\left |p_x\right>+i\left |p_y\right>)/\sqrt{2}$, $\left |3\right>=(\left |p_x\right>-i\left |p_y\right>)/\sqrt{2}$. 

The following set of parameters describing a quantum emitter is used in this paper: the transition dipole moment is $10$ Debye and the radiationless lifetime of the excited state is 1 ps. The number density and the pure dephasing time are varied.

\section{Results and discussion}

The intent of this paper is to scrutinize the optical properties of a periodic system comprised of a periodic array of V-grooves in an optically thick silver film (the thickness of the film in all simulations is 800 nm) that is optically coupled to quantum emitters. We first consider a periodic array of V-grooves without emitters, followed by a hybrid system consisting of emitters (all starting in the ground state) added into the grooves, and finally this same system with an optical femtosecond pump applied.  

\begin{figure}
\begin{center}
\includegraphics[width=0.9\textwidth]{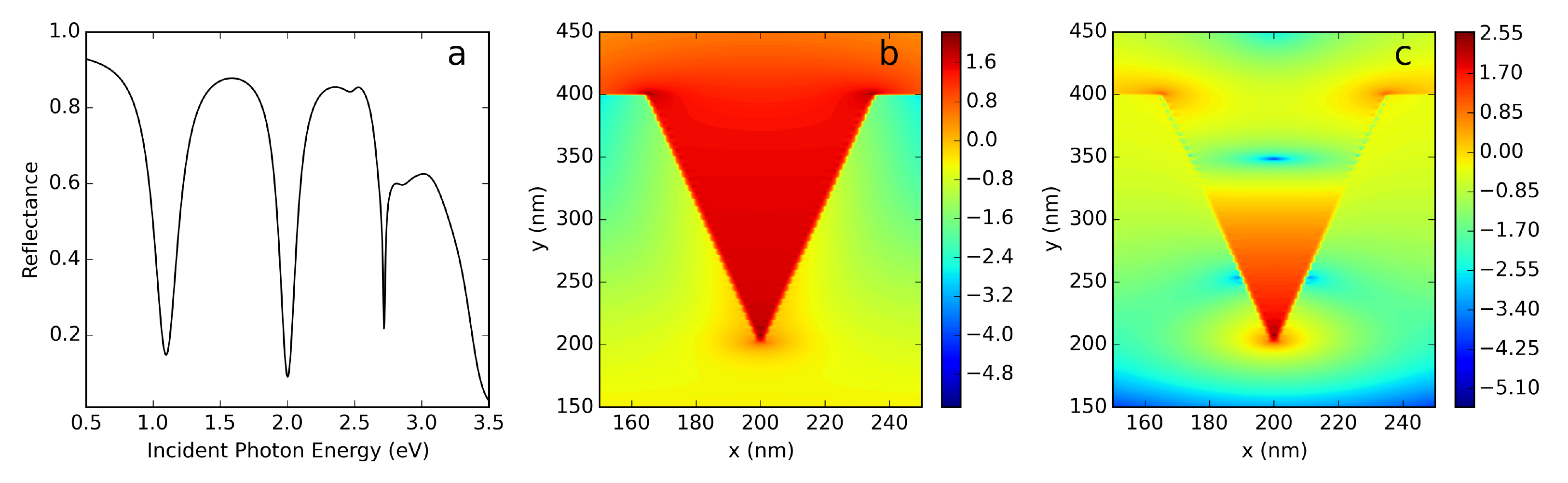}
\caption{\label{figure2} Linear optical response of bare V-grooves at normal incidence. Panel (a) shows reflection spectrum of bare grooves with $400$ nm period, $20$ degree groove angle and $200$ nm groove depth. Panel (b) shows time-averaged intensity in the bare groove when excited by CW plane wave at $1.2$ eV. The intensity is distributed throughout the groove. Panel (c) shows time-averaged intensity in the bare groove when excited with CW plane wave at $2.255$ eV. The intensity is localized near the bottom of the groove. The contour data in panels (b) and (c) is logarithmic and normalized to the incident intensity.}
\end{center}
\end{figure}

The reflection spectrum of a bare silver grating with a $400$ nm period obtained at normal incidence is shown in Fig. \ref{figure2}a. Three well-resolved resonances are observed with the energy of each depending on the geometrical parameters of the grating. Both of the resonances at lower energy are thought to be of a plasmonic, rather than geometrical, nature as each disappears when the conductivity of the metal is made infinite (i.e. perfectly reflecting). The resonance with the highest energy is a Wood's anomaly, which is indicated by its wavelength corresponding to the groove period. Each of these resonances is observed in [\onlinecite{Sondergaard2009}], where their behavior was analyzed by adjusting the geometry of the V-grooves as well as the angle of incidence. The incident fields in our simulations are $p$-polarized with normal incidence, as $s$-polarized waves cannot excite SPP waves. In [\onlinecite{Sondergaard2009}], the field enhancements within the groove are substantially smaller for $s$-polarized waves than $p$-polarized waves.

As noted in [\onlinecite{Sondergaard2009}], the intensity of the lower energy resonance is distributed throughout the groove whereas that of the higher energy resonance is localized near the bottom (see Fig. \ref{figure2}). 

On account of each resonance being well-defined for this geometry, the remainder of this work (unless otherwise specified) will focus on V-grooves with a period of $400$ nm, a groove angle of $20$ degrees, a groove depth of $200$ nm, and normally incident fields.

\begin{figure}[t!]
\begin{center}
\includegraphics[width=0.8\textwidth]{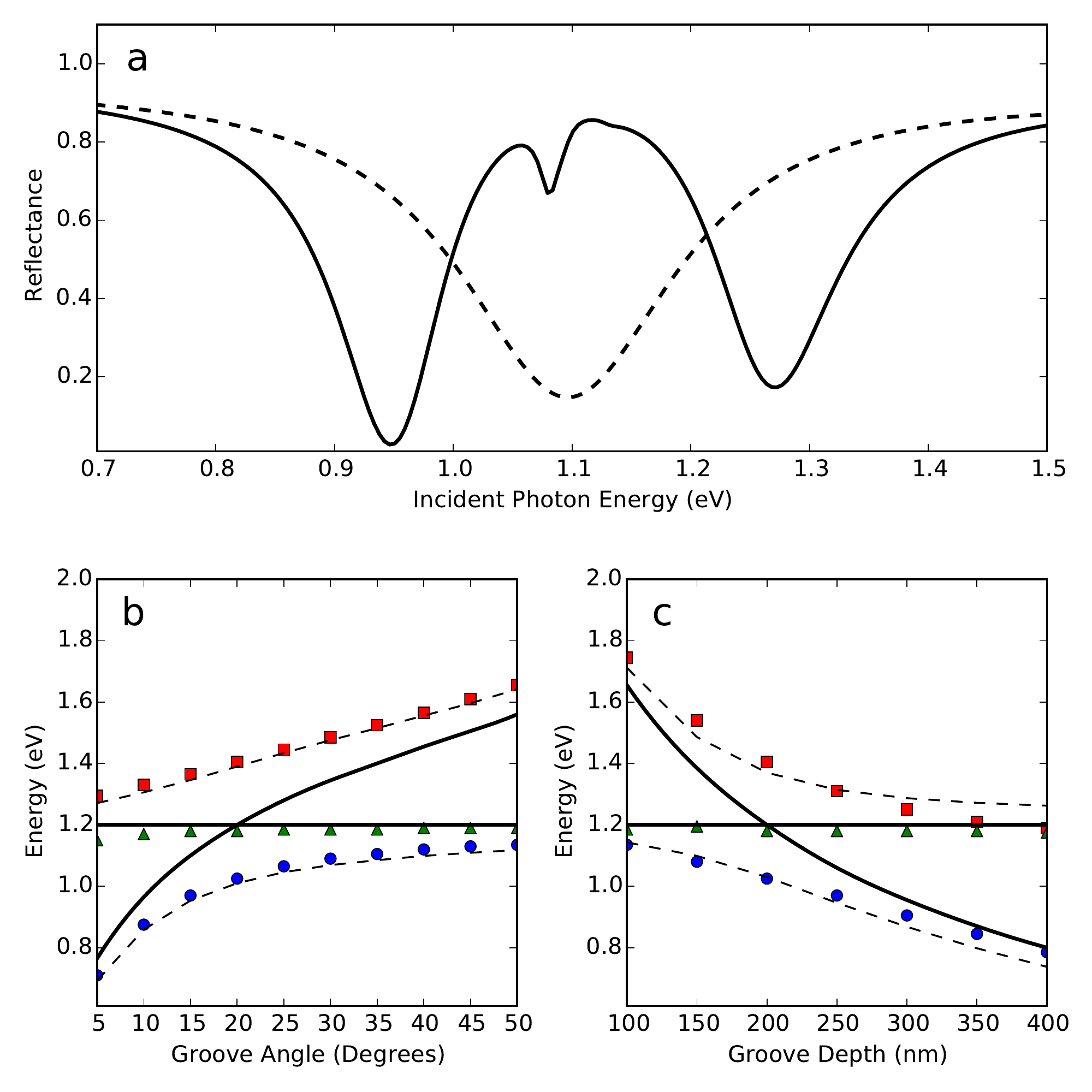}
\caption{\label{figure3} Optics of periodic V-grooves coupled to quantum emitters. Panel (a) shows the reflection for bare grooves (dashed line) and the reflection for grooves with emitters is shown as a solid line (with emitters resonant at 1.2 eV). Panel (b) shows the upper polariton (red squares), lower polariton (blue circles), and third resonant mode (green triangles) as a function of the groove angle to sweep the SPP resonance through the emitter resonance. Panel (c) shows the same modes as in panel (b) but as functions of the groove depth to sweep the SPP resonance through the emitter resonance. The horizontal black lines in panels (b) and (c) represent the fixed emitter resonance, the curved black lines represent the SPP energy of the bare grooves for the given geometry, and the dashed lines are values predicted by the coupled oscillator model. The density of emitters is $3\times10^{26}$ emitters/$\text{m}^{3}$.}
\end{center}
\end{figure}

We now consider the optical response of the system when emitters are added inside the grooves. When the emitters are resonant to the structure, normal mode splitting (Rabi splitting) is clearly observed (Fig. \ref{figure3}a). The Rabi splitting reaches $325$ meV. This amount of splitting is considered large for hybrid nanostructures \cite{Schlather2013}. Either at high emitter densities or a large transition dipole moment a third feature appears at or near the emitter resonance as  can be seen in Fig. \ref{figure3}a near $1.1$ eV. This feature, which is not predicted by the coupled-oscillator model, has been observed previously in simulations \cite{Salomon2012, Sukharev2013, doi:10.1021/ph500032d} as well as in experiments \cite{Hutchison2011, ANIE:ANIE200903191, Sugawara2006}. Several observations discussed in [\onlinecite{Salomon2012}] suggest that this peak has its origins in SPP enhanced emitter-emitter interactions. 

To better understand the physics of this resonance we performed simulations gradually varying either groove angle (Fig. \ref{figure3}b) or groove depth (Fig. \ref{figure3}c). Note that the bare SPP lines plotted in Fig. \ref{figure3}b and \ref{figure3}c are clearly not linear with groove angle or groove depth (see below). This allows us to sweep the SPP resonance through the emitter's mode. In the reflection spectrum we record energy positions of lower and upper branch of the hybrid mode. This is carried out at the emitter density of $3\times10^{26}$ emitters/$\text{m}^3$ (at which the third feature is prominent) and the results are shown in Fig. \ref{figure3}b and  \ref{figure3}c. The results clearly indicate avoided crossing - a unique signature of the strong coupling due to efficient energy exchange between the corresponding SPP mode and molecular excitons (in our case these are simply two-level emitters). Next, the energy of the intermediate peak (green triangles in Fig. \ref{figure3}b and c) does not deviate appreciably from the emitter resonance even as the SPP resonance is tuned. Furthermore, this peak merges with the upper polariton as the thickness of a spacer layer between the grating and emitters increases. The dipole coupling between the emitters themselves is therefore suspected given the fall-off of this peak as the SPP field at the emitters' location decreases. Finally, a simulation was run in which the entire region containing emitters was replaced by a single two-level system (essentially a spatially distributed single emitter with a very large dipole moment), thereby eliminating any possible interaction between the emitters. In this case, the third peak disappears even under extremely high coupling conditions. This confirms earlier findings \cite{Salomon2012, doi:10.1021/ph500032d} which suggested that the intermediate resonance located in the middle of the Rabi splitting corresponds to dipole-dipole interactions between emitters greatly enhanced by the SPP mode.

The dashed lines in Fig. \ref{figure3}b and \ref{figure3}c are calculations of the values of the upper and lower polaritons using the coupled oscillator model. The Hamiltonian of the coupled system is
\begin{equation}
\begin{bmatrix} 
  E_{m} &  \Delta\\ 
  \Delta & E_{pl} 
\end{bmatrix}
\end{equation}
$E_{pl}$ is the SPP energy of the bare metallic grooves (obtained from simulations), $E_m$ is the transition energy of uncoupled emitters, and $2\Delta$ is the minimum Rabi splitting value. The eigenvalues obtained are
\begin{equation}
E_{U,L} = \frac{(E_{pl}+E_m) \pm \sqrt{(E_{pl}+E_m)^{2} - 4\Delta^{2}}}{2}
\end{equation} 
$E_{U,L}$ are the energies of the upper or lower polaritons. In Fig. \ref{figure3}b, the comparison to the coupled oscillator model is close whereas in Fig. \ref{figure3}c it deviates somewhat at larger groove depths. In particular, we note that both the upper and lower polaritons appear to be "pinched" toward each other in Fig. \ref{figure3}c. We offer two possible explanations for this. First, a large groove depth requires the incident fields to traverse a larger length of emitters. More absorption takes place than for a shallower grove, leading to decreased excitation of SPP waves near the bottom and therefore less coupling. Second, the third feature may interact with the upper and lower polaritons at greater groove depths in such a way as to reduce the coupling.   

\begin{figure}[t!]
\begin{center}
\includegraphics[width=1.0\textwidth]{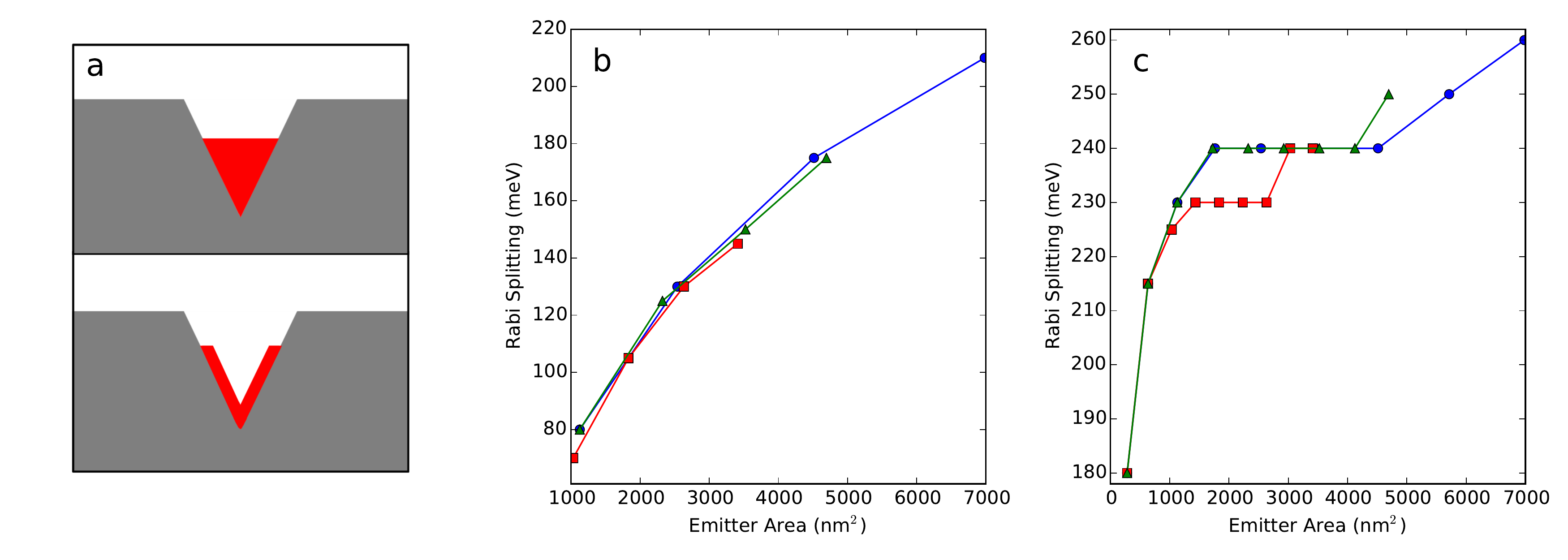}
\caption{\label{figure4} Spatially dependent coupling. The density of emitters is $10^{26}$ emitters/$\text{m}^{3}$, the transition energy is $1.2$ eV, and the pure dephasing time is $400$ fs. As the height of the molecular aggregate is increased, the coupling for the distributed resonance increases continuously whereas that for the localized resonance levels off quickly. Panel (a) shows schematics of "full" grooves versus "hollowed" grooves. Panel (b) shows the Rabi splitting as a function of the area occupied by the emitters for the more spatially distributed $1.2$ eV resonance, where blue circles indicate full grooves, red squares indicate hollowed grooves (width of emitter region is $10 $nm) and green triangles indicate hollowed grooves (width of emitter region is $15$ nm). Panel (c) is the same as panel (b) except the data is shown for the more localized $2.255$ eV resonance. The pure dephasing time is $600$ fs.} 
\end{center}
\end{figure}

While it is clear that many of the emitters are coupled to the SPPs, some may remain coupled only to the incident field \cite{Agranovich2005} and thus act as an absorbing layer. To better understand the overall optical coupling in spatially distributed inhomogeneous hybrid systems, two different arrangements of emitters are simulated: "full" grooves and "hollowed" grooves, as shown in Fig. \ref{figure4}a. In the "full" grooves, the region of the grooves occupied by emitters is completely full of emitters up to a given height, whereas in "hollowed" grooves, the region of emitters extends out sideways from either side of the groove by a fixed width (referred to as "width of emitters" from here on). The grooves are illuminated with CW fields at $1.2$ eV (uniform intensity, Fig. \ref{figure2}b) or $2.255$ eV (localized intensity, Fig. \ref{figure2}c). 

The Rabi splitting is observed as the area occupied by the emitters is varied. Fig. \ref{figure4}b shows that, for the distributed resonance, there is relatively little difference in Rabi splitting when the same amount of area is occupied by the emitters for either full or hollowed grooves, which demonstrates that the entire volume of emitters is indeed coupled to the SPP waves. Given that coupling strength depends on field strength, this is reasonable in light of the uniform distribution of intensity in the bare groove at this frequency; the same Rabi splitting is achieved regardless of where a given area of emitters is placed in the groove. Fig. \ref{figure4}c shows that, for the localized resonance, a large increase in Rabi splitting occurs as the area is increased from the bottom, but then it levels off to a constant value well before the emitters reach the top of the groove. This is exactly what is expected as the strong fields near the bottom of the groove give the strongest coupling. The constant value of Rabi splitting is smaller for the lowest width of emitters (10 nm) and this occurs because the fields above the bottom of the groove are weaker but not zero (hence a larger hollow region loses some of the emitters coupled to those weaker fields). A small jump occurs at the end of each graph in Fig. \ref{figure4}c and this is due to the increased fields at the sharp corners of the groove. This jump occurs at smaller areas for more hollowed grooves as they occupy less area when they extend to the top of the groove. 

All simulations discussed above begin with all of the emitters in the ground state. One can pump the system by sending in a high-intensity pulse, thereby inducing Rabi oscillations in the emitters. Below the time dynamics of a pumped hybrid system is discussed. In particular, we observe how Rabi splitting depends on the ground state population at the end of the pump.

\begin{figure}[t!]
\begin{center}
\includegraphics[width=0.9\textwidth]{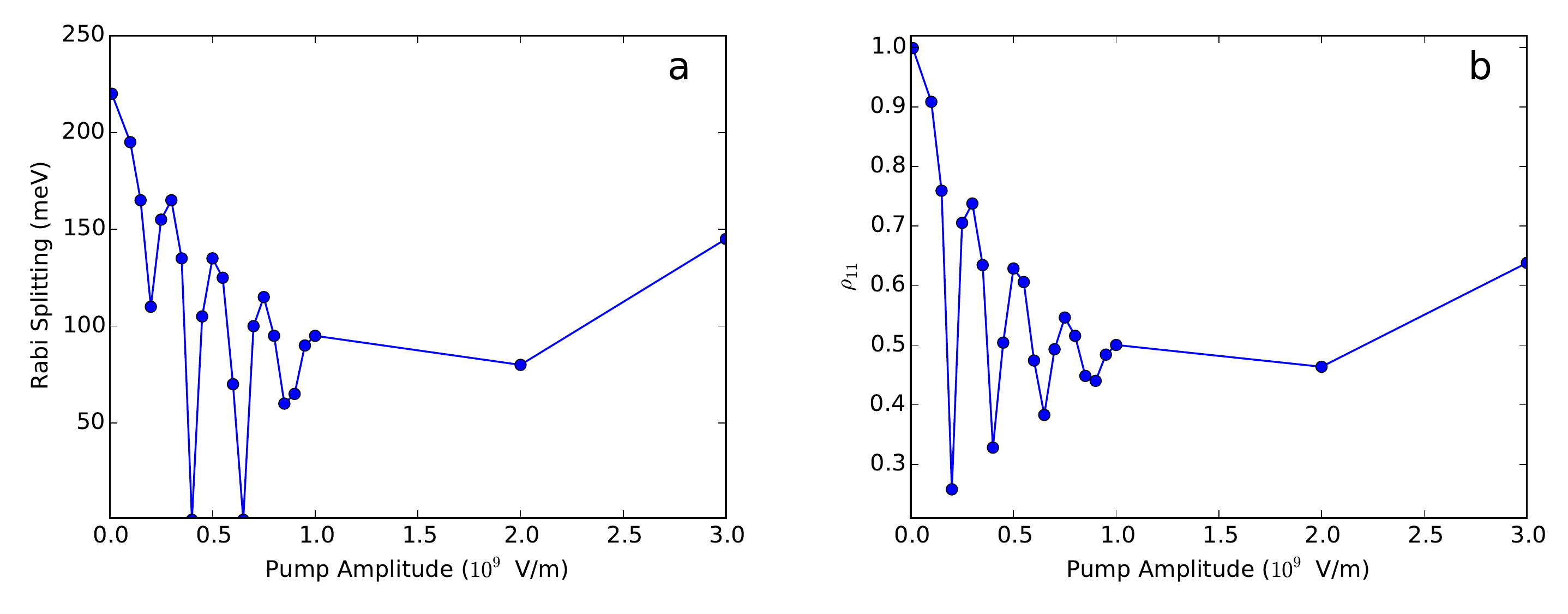}
\caption{\label{figure5} Pump-probe dynamics. Panel (a) shows Rabi splitting as a function of pump amplitude. Each value is obtained by launching a short, low-intensity probe pulse immediately after the pump. Panel (b) shows ground state population, $\rho_{11}$, as a function of pump amplitude. This value is averaged over the entire region occupied by emitters and obtained at the end of the pump. The density of emitters is $10^{26}$ emitters/$\text{m}^{3}$, the pure dephasing time is 400 fs, the transition energy is $1.2$ eV, and the duration of the pump is $30$ fs.}
\end{center}
\end{figure}

First, the system is pumped with a $30$ fs pulse, and it is subsequently probed with a short, low-intensity pulse as it was in the linear regime. The Rabi splitting is obtained from the reflection spectrum by calculating the difference in resonant energies for the upper and lower polaritons. Fig. \ref{figure5}a shows Rabi splitting as a function of pump amplitude at the end of the pump. Fig. \ref{figure5}b shows the ground state population, averaged over the entire region of emitters at the end of the pump, as a function of pump amplitude. Notice how the plot in Fig. \ref{figure5}b oscillates; this is because the area under the pump pulse (which depends on both pump amplitude and duration) determines the number of Rabi cycles \cite{Eberly}. Thus different pump amplitudes generate greater or fewer Rabi oscillations, leading to different values of the ground state population at the end of the pump. 

Fig. \ref{figure5} indicates that the amount of Rabi splitting depends on the ground state population of the emitters, $\rho_{11}$. We see that a large excited population of emitters in the entire groove will reduce or eliminate the Rabi splitting whereas a smaller excited population will yield larger Rabi splitting. We speculate that the coupling may take on a different character in the non-linear regime, perhaps to the extent that Rabi splitting is not the only indicator of coupling strength. Further investigation of the optical properties of excited emitters coupled to plasmons is clearly warranted. 

Although the changes of $\rho_{11}$ are in step with changes in Rabi splitting, large decreases of $\rho_{11}$ do not always give a correspondingly large decrease in Rabi splitting and this can be understood in terms of the non-uniformity of $\rho_{11}$ throughout the groove. In particular, the pumping fields arrive at deeper parts of the groove later than they arrive at the top and the Rabi oscillations are not perfectly in phase with one another along the length of the groove due to retardation. Additionally, the SPP fields are inhomogeneous throughout the groove. Thus, the spatially averaged value of $\rho_{11}$ after the pump gives a good, though not ideal, indication of the subsequent coupling to SPP fields.

\begin{figure}[t!]
\begin{center}
\includegraphics[width=0.8\textwidth]{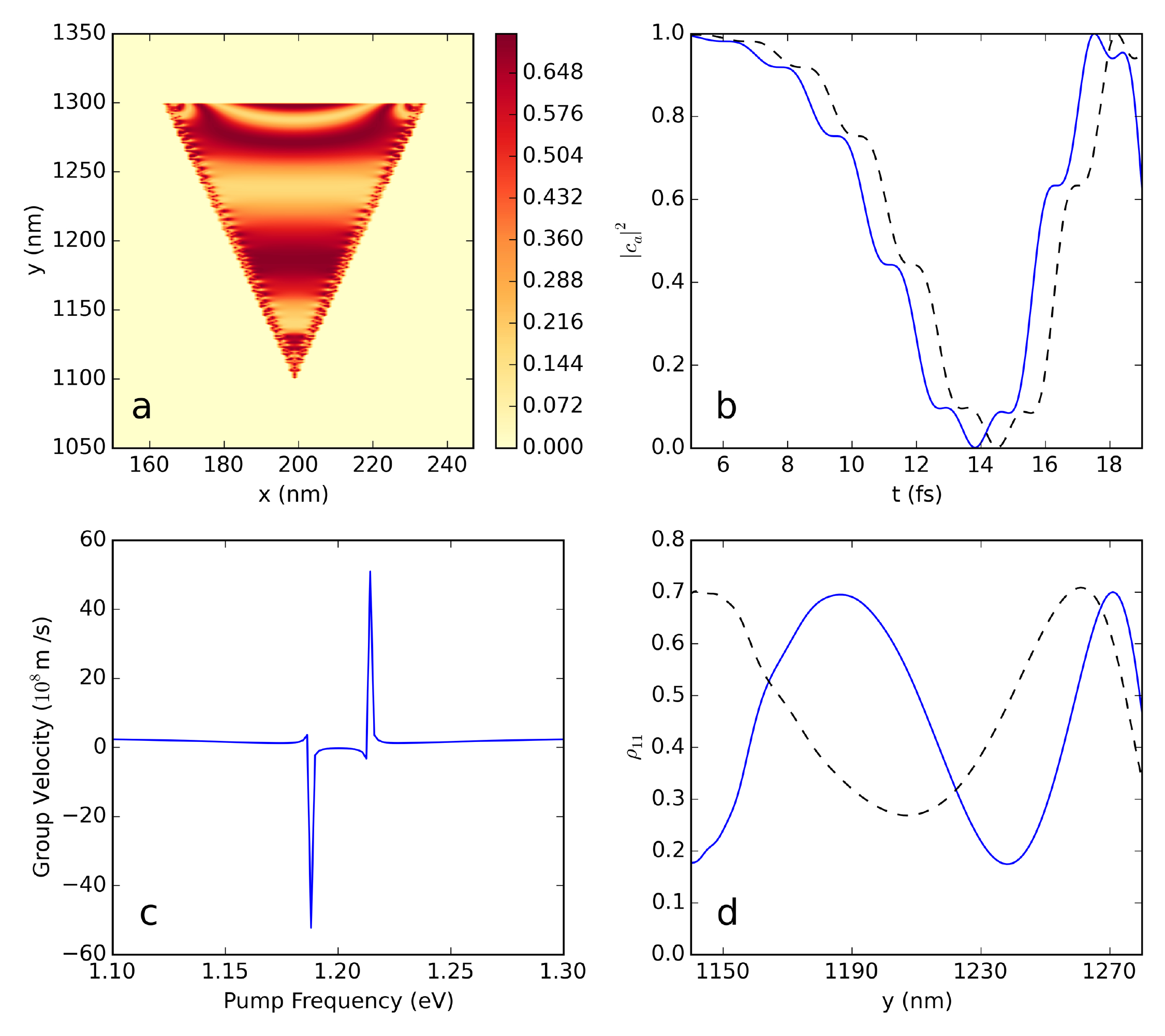}
\caption{\label{figure6} Spatial modulations of the molecular ground state. Each run consists of a $70$ fs pump applied to the hybrid V-grooves system. The simulations shown in panels (a), (c), and (d) are run with FDTD and the Liouville-von Neumann equation whereas that in panel (b) is run by numerically integrating the Schrodinger equation for a two-level atom. For panels (a), (c), and (d) the density is $10^{26}$ emitters$/m^{3}$, the pump amplitude is $7\times 10^{8}$ V/m, and the dephasing time is $400$ fs. Panel (a) shows the fraction of emitters in the ground state. The characteristic length of spatial modulations is much less than the pump wavelength of $1033$ nm. Panel (b) shows oscillations of the ground state population for two emitters separated by $50$ nm. Panel (c) shows that the group velocity at the transition frequency is less than $c$. Panel (d) shows the ground state population as a function of coordinate for the system pumped on resonance ($1.2$ eV, solid line) and slightly off resonance ($1.15$ eV, dashed line).} 
\end{center}
\end{figure} 

Fig. \ref{figure6}a shows a spatial distribution of $\rho_{11}$ at the end of $70$ fs long pump. A clear strong spatial variation of the ground state population is seen. In fact, the variations of the ground state population are oscillations whose wavelength varies somewhat over the region of emitters. One might surmise that the wavelength of these oscillations is on the order of that of the pump, but it is actually significantly smaller. To investigate this further, we numerically solve the Schr\"{o}dinger equation for a one-dimensional region of two-level atoms (finite along, say, the $z$-axis while infinite along two others) subject to excitation from a pump.

Because we consider a one-dimensional region, retardation effects must be included by using the retarded time $t - z/v$, where v is the speed of light in the medium in the expression for the pulse. The retardation effects are revealed to be the cause of the spatial modulations of the ground state population, as shown in Fig. \ref{figure6}b: the temporal oscillations of the ground state probability are similar between adjacent spatial points, but slightly shifted. Hence at a given time, adjacent points have slightly different values of ground state probability. To further elucidate this idea, we generated larger phase shifts in the temporal oscillations between nearby points by adjusting two parameters: the pump amplitude and the propagation velocity of light in the emitter region. 

For a larger pump amplitude, each emitter will undergo Rabi flopping more rapidly in time. Thus a given phase shift between adjacent locations will lead to a larger shift in the ground state population between those two locations. This was observed in our simulations: in general, increasing the pump amplitude leads to an increase in the number of spatial oscillations of the ground state population over the region of emitters. For a slower group velocity, the pump takes longer to reach an adjacent point, causing the temporal oscillations between two adjacent points to acquire a larger phase difference. Our simulations allow for the adjustment of the group velocity of light in the emitter region, and we see that a decrease in the group velocity results in a greater number of spatial oscillations relative to a larger speed of light.   

Fig. \ref{figure6}c shows that group velocity is decreased (in fact, negative since most emitters are inverted) at the transition frequency of $1.2$ eV, and it is higher away from resonance. A negative group velocity is possible in inverted systems and has been experimentally observed \cite{Wang2000}. It is stated that this occurs when different frequency components of a pulse interfere with one another (in a region of anomalous dispersion) in such a way as to cause a resonant pulse to be advanced relative to a non-resonant pulse traveling at $c$. We are assured by [\onlinecite{Siddiqui2003}] that this does not violate causality and that it occurs because the early parts of the pulse are reshaped to resemble the later pulse. Furthermore, [\onlinecite{Mojahedi2003}] points out that in passing through a medium with negative group velocity, the information transmitted by the pulse front \textit{suffers a positive and causal delay}. Our system was pumped (in separate runs) both on and off of resonance and Fig. \ref{figure6}d shows that the wavelength of the spatial oscillations for the \textit{off resonance} pump is indeed increased relative to the \textit{on resonance} pump on account of the latter having a slower group velocity.

All of these results indicate that retardation effects give rise to spatial oscillations in the ground state population of the emitters contained within the groove. The larger fields of the surface plasmons and the slower group velocity within the emitter region surely enhance these spatial oscillations. It may be possible to use this phenomenon to perform a new type of optical design using femtosecond pulses as a tool to craft hybrid systems. Highly inhomogeneous spatial modulations of molecules lead to the modified refractive index, which is appreciably anisotropic. One can envision an exciting opportunity for a new research direction, in which both the geometry of metal nanostructures and pump pulses govern the refractive index of the system. Furthermore one can apply optimization techniques such as genetic algorithms \cite{Brixner:2003aa}, for instance, to design materials with desired optical properties.

\section{Conclusion}
\label{sec:Conclusion}
The optical properties of the bare and hybrid V-groove systems have been explored under several different circumstances. Two SPP resonances with entirely different spatial distributions were shown. Optical coupling between quantum emitters and SPP waves was thoroughly characterized by simulating different groove geometries and spatial configurations of emitters' distributions within the grooves. The value of Rabi splitting was shown to vary with the pumping intensity and in accord with the ground state population. An explanation of this awaits a more developed understanding of coupling in the non-linear regime. Spatial oscillations of the ground state population of emitters within the groove are shown to be the result of field retardation. This work puts forth several suggestions for experimental investigation of coupling in hybrid systems, most notably an investigation of the time dynamics of coupling when the system is pumped with different intensities.

\section*{Acknowledgements}
  The authors acknowledge support from the Air Force Office of Scientific Research (Summer Faculty Fellowship 2013).


\begin{thebibliography}{38}%
\makeatletter
\providecommand \@ifxundefined [1]{%
 \@ifx{#1\undefined}
}%
\providecommand \@ifnum [1]{%
 \ifnum #1\expandafter \@firstoftwo
 \else \expandafter \@secondoftwo
 \fi
}%
\providecommand \@ifx [1]{%
 \ifx #1\expandafter \@firstoftwo
 \else \expandafter \@secondoftwo
 \fi
}%
\providecommand \natexlab [1]{#1}%
\providecommand \enquote  [1]{``#1''}%
\providecommand \bibnamefont  [1]{#1}%
\providecommand \bibfnamefont [1]{#1}%
\providecommand \citenamefont [1]{#1}%
\providecommand \href@noop [0]{\@secondoftwo}%
\providecommand \href [0]{\begingroup \@sanitize@url \@href}%
\providecommand \@href[1]{\@@startlink{#1}\@@href}%
\providecommand \@@href[1]{\endgroup#1\@@endlink}%
\providecommand \@sanitize@url [0]{\catcode `\\12\catcode `\$12\catcode
  `\&12\catcode `\#12\catcode `\^12\catcode `\_12\catcode `\%12\relax}%
\providecommand \@@startlink[1]{}%
\providecommand \@@endlink[0]{}%
\providecommand \url  [0]{\begingroup\@sanitize@url \@url }%
\providecommand \@url [1]{\endgroup\@href {#1}{\urlprefix }}%
\providecommand \urlprefix  [0]{URL }%
\providecommand \Eprint [0]{\href }%
\providecommand \doibase [0]{http://dx.doi.org/}%
\providecommand \selectlanguage [0]{\@gobble}%
\providecommand \bibinfo  [0]{\@secondoftwo}%
\providecommand \bibfield  [0]{\@secondoftwo}%
\providecommand \translation [1]{[#1]}%
\providecommand \BibitemOpen [0]{}%
\providecommand \bibitemStop [0]{}%
\providecommand \bibitemNoStop [0]{.\EOS\space}%
\providecommand \EOS [0]{\spacefactor3000\relax}%
\providecommand \BibitemShut  [1]{\csname bibitem#1\endcsname}%
\let\auto@bib@innerbib\@empty
\bibitem [{\citenamefont {Zayats}\ \emph {et~al.}(2005)\citenamefont {Zayats},
  \citenamefont {Smolyaninov},\ and\ \citenamefont
  {Maradudin}}]{Zayats2005131}%
  \BibitemOpen
  \bibfield  {author} {\bibinfo {author} {\bibfnamefont {A.~V.}\ \bibnamefont
  {Zayats}}, \bibinfo {author} {\bibfnamefont {I.~I.}\ \bibnamefont
  {Smolyaninov}}, \ and\ \bibinfo {author} {\bibfnamefont {A.~A.}\ \bibnamefont
  {Maradudin}},\ }\href {\doibase
  http://dx.doi.org/10.1016/j.physrep.2004.11.001} {\bibfield  {journal}
  {\bibinfo  {journal} {Phys Rep}\ }\textbf {\bibinfo {volume} {408}},\
  \bibinfo {pages} {131 } (\bibinfo {year} {2005})}\BibitemShut {NoStop}%
\bibitem [{\citenamefont {Murray}\ and\ \citenamefont
  {Barnes}(2007)}]{ADMA:ADMA200700678}%
  \BibitemOpen
  \bibfield  {author} {\bibinfo {author} {\bibfnamefont {W.~A.}\ \bibnamefont
  {Murray}}\ and\ \bibinfo {author} {\bibfnamefont {W.~L.}\ \bibnamefont
  {Barnes}},\ }\href {\doibase 10.1002/adma.200700678} {\bibfield  {journal}
  {\bibinfo  {journal} {Adv. Mater.}\ }\textbf {\bibinfo {volume} {19}},\
  \bibinfo {pages} {3771} (\bibinfo {year} {2007})}\BibitemShut {NoStop}%
\bibitem [{\citenamefont {Han}\ and\ \citenamefont
  {Bozhevolnyi}(2013)}]{Han:2013aa}%
  \BibitemOpen
  \bibfield  {author} {\bibinfo {author} {\bibfnamefont {Z.}~\bibnamefont
  {Han}}\ and\ \bibinfo {author} {\bibfnamefont {S.~I.}\ \bibnamefont
  {Bozhevolnyi}},\ }\href {http://stacks.iop.org/0034-4885/76/i=1/a=016402}
  {\bibfield  {journal} {\bibinfo  {journal} {Rep. Prog. Phys.}\ }\textbf
  {\bibinfo {volume} {76}},\ \bibinfo {pages} {016402} (\bibinfo {year}
  {2013})}\BibitemShut {NoStop}%
\bibitem [{\citenamefont {Gramotnev}\ and\ \citenamefont
  {Bozhevolnyi}(2014)}]{Gramotnev:2014aa}%
  \BibitemOpen
  \bibfield  {author} {\bibinfo {author} {\bibfnamefont {D.~K.}\ \bibnamefont
  {Gramotnev}}\ and\ \bibinfo {author} {\bibfnamefont {S.~I.}\ \bibnamefont
  {Bozhevolnyi}},\ }\href {http://dx.doi.org/10.1038/nphoton.2013.232}
  {\bibfield  {journal} {\bibinfo  {journal} {Nat. Photon.}\ }\textbf {\bibinfo
  {volume} {8}},\ \bibinfo {pages} {13} (\bibinfo {year} {2014})}\BibitemShut
  {NoStop}%
\bibitem [{\citenamefont {Noginov}\ \emph {et~al.}(2009)\citenamefont
  {Noginov}, \citenamefont {Zhu}, \citenamefont {Belgrave}, \citenamefont
  {Bakker}, \citenamefont {Shalaev}, \citenamefont {Narimanov}, \citenamefont
  {Stout}, \citenamefont {Herz}, \citenamefont {Suteewong},\ and\ \citenamefont
  {Wiesner}}]{Noginov:2009aa}%
  \BibitemOpen
  \bibfield  {author} {\bibinfo {author} {\bibfnamefont {M.~A.}\ \bibnamefont
  {Noginov}}, \bibinfo {author} {\bibfnamefont {G.}~\bibnamefont {Zhu}},
  \bibinfo {author} {\bibfnamefont {A.~M.}\ \bibnamefont {Belgrave}}, \bibinfo
  {author} {\bibfnamefont {R.}~\bibnamefont {Bakker}}, \bibinfo {author}
  {\bibfnamefont {V.~M.}\ \bibnamefont {Shalaev}}, \bibinfo {author}
  {\bibfnamefont {E.~E.}\ \bibnamefont {Narimanov}}, \bibinfo {author}
  {\bibfnamefont {S.}~\bibnamefont {Stout}}, \bibinfo {author} {\bibfnamefont
  {E.}~\bibnamefont {Herz}}, \bibinfo {author} {\bibfnamefont {T.}~\bibnamefont
  {Suteewong}}, \ and\ \bibinfo {author} {\bibfnamefont {U.}~\bibnamefont
  {Wiesner}},\ }\href {http://dx.doi.org/10.1038/nature08318} {\bibfield
  {journal} {\bibinfo  {journal} {Nature}\ }\textbf {\bibinfo {volume} {460}},\
  \bibinfo {pages} {1110} (\bibinfo {year} {2009})}\BibitemShut {NoStop}%
\bibitem [{\citenamefont {Stockman}(2011)}]{Stockman:11}%
  \BibitemOpen
  \bibfield  {author} {\bibinfo {author} {\bibfnamefont {M.~I.}\ \bibnamefont
  {Stockman}},\ }\href {\doibase 10.1364/OE.19.022029} {\bibfield  {journal}
  {\bibinfo  {journal} {Opt. Express}\ }\textbf {\bibinfo {volume} {19}},\
  \bibinfo {pages} {22029} (\bibinfo {year} {2011})}\BibitemShut {NoStop}%
\bibitem [{\citenamefont {Torma}\ and\ \citenamefont {Barnes}(2015)}]{Torma}%
  \BibitemOpen
  \bibfield  {author} {\bibinfo {author} {\bibfnamefont {P.}~\bibnamefont
  {Torma}}\ and\ \bibinfo {author} {\bibfnamefont {W.~L.}\ \bibnamefont
  {Barnes}},\ }\href {http://stacks.iop.org/0034-4885/78/i=1/a=013901}
  {\bibfield  {journal} {\bibinfo  {journal} {Rep. Prog. Phys.}\ }\textbf
  {\bibinfo {volume} {78}},\ \bibinfo {pages} {013901} (\bibinfo {year}
  {2015})}\BibitemShut {NoStop}%
\bibitem [{\citenamefont {Salomon}\ \emph
  {et~al.}(2009{\natexlab{a}})\citenamefont {Salomon}, \citenamefont {Genet},\
  and\ \citenamefont {Ebbesen}}]{ANIE:ANIE200903191}%
  \BibitemOpen
  \bibfield  {author} {\bibinfo {author} {\bibfnamefont {A.}~\bibnamefont
  {Salomon}}, \bibinfo {author} {\bibfnamefont {C.}~\bibnamefont {Genet}}, \
  and\ \bibinfo {author} {\bibfnamefont {T.}~\bibnamefont {Ebbesen}},\ }\href
  {\doibase 10.1002/anie.200903191} {\bibfield  {journal} {\bibinfo  {journal}
  {Angew. Chem. Int. Edit.}\ }\textbf {\bibinfo {volume} {48}},\ \bibinfo
  {pages} {8748} (\bibinfo {year} {2009}{\natexlab{a}})}\BibitemShut {NoStop}%
\bibitem [{\citenamefont {Vasa}\ \emph {et~al.}(2013)\citenamefont {Vasa},
  \citenamefont {Wang}, \citenamefont {Pomraenke}, \citenamefont {Lammers},
  \citenamefont {Maiuri}, \citenamefont {Manzoni}, \citenamefont {Cerullo},\
  and\ \citenamefont {Lienau}}]{Vasa2013}%
  \BibitemOpen
  \bibfield  {author} {\bibinfo {author} {\bibfnamefont {P.}~\bibnamefont
  {Vasa}}, \bibinfo {author} {\bibfnamefont {W.}~\bibnamefont {Wang}}, \bibinfo
  {author} {\bibfnamefont {R.}~\bibnamefont {Pomraenke}}, \bibinfo {author}
  {\bibfnamefont {M.}~\bibnamefont {Lammers}}, \bibinfo {author} {\bibfnamefont
  {M.}~\bibnamefont {Maiuri}}, \bibinfo {author} {\bibfnamefont
  {C.}~\bibnamefont {Manzoni}}, \bibinfo {author} {\bibfnamefont
  {G.}~\bibnamefont {Cerullo}}, \ and\ \bibinfo {author} {\bibfnamefont
  {C.}~\bibnamefont {Lienau}},\ }\href
  {http://dx.doi.org/10.1038/nphoton.2012.340} {\bibfield  {journal} {\bibinfo
  {journal} {Nat Photon}\ }\textbf {\bibinfo {volume} {7}},\ \bibinfo {pages}
  {128} (\bibinfo {year} {2013})}\BibitemShut {NoStop}%
\bibitem [{\citenamefont {Salomon}\ \emph
  {et~al.}(2009{\natexlab{b}})\citenamefont {Salomon}, \citenamefont {Genet},\
  and\ \citenamefont {Ebbesen}}]{Salomon2009}%
  \BibitemOpen
  \bibfield  {author} {\bibinfo {author} {\bibfnamefont {A.}~\bibnamefont
  {Salomon}}, \bibinfo {author} {\bibfnamefont {C.}~\bibnamefont {Genet}}, \
  and\ \bibinfo {author} {\bibfnamefont {T.}~\bibnamefont {Ebbesen}},\ }\href
  {\doibase 10.1002/anie.200903191} {\bibfield  {journal} {\bibinfo  {journal}
  {Angew Chem Int Edit}\ }\textbf {\bibinfo {volume} {48}},\ \bibinfo {pages}
  {8748} (\bibinfo {year} {2009}{\natexlab{b}})}\BibitemShut {NoStop}%
\bibitem [{\citenamefont {Symonds}\ \emph {et~al.}(2008)\citenamefont
  {Symonds}, \citenamefont {Bonnand}, \citenamefont {Plenet}, \citenamefont
  {Br{\'e}hier}, \citenamefont {Parashkov}, \citenamefont {Lauret},
  \citenamefont {Deleporte},\ and\ \citenamefont
  {Bellessa}}]{1367-2630-10-6-065017}%
  \BibitemOpen
  \bibfield  {author} {\bibinfo {author} {\bibfnamefont {C.}~\bibnamefont
  {Symonds}}, \bibinfo {author} {\bibfnamefont {C.}~\bibnamefont {Bonnand}},
  \bibinfo {author} {\bibfnamefont {J.~C.}\ \bibnamefont {Plenet}}, \bibinfo
  {author} {\bibfnamefont {A.}~\bibnamefont {Br{\'e}hier}}, \bibinfo {author}
  {\bibfnamefont {R.}~\bibnamefont {Parashkov}}, \bibinfo {author}
  {\bibfnamefont {J.~S.}\ \bibnamefont {Lauret}}, \bibinfo {author}
  {\bibfnamefont {E.}~\bibnamefont {Deleporte}}, \ and\ \bibinfo {author}
  {\bibfnamefont {J.}~\bibnamefont {Bellessa}},\ }\href
  {http://stacks.iop.org/1367-2630/10/i=6/a=065017} {\bibfield  {journal}
  {\bibinfo  {journal} {New J. Phys.}\ }\textbf {\bibinfo {volume} {10}},\
  \bibinfo {pages} {065017} (\bibinfo {year} {2008})}\BibitemShut {NoStop}%
\bibitem [{\citenamefont {Vasa}\ \emph {et~al.}(2010)\citenamefont {Vasa},
  \citenamefont {Pomraenke}, \citenamefont {Cirmi}, \citenamefont {De~Re},
  \citenamefont {Wang}, \citenamefont {Schwieger}, \citenamefont {Leipold},
  \citenamefont {Runge}, \citenamefont {Cerullo},\ and\ \citenamefont
  {Lienau}}]{Vasa2010}%
  \BibitemOpen
  \bibfield  {author} {\bibinfo {author} {\bibfnamefont {P.}~\bibnamefont
  {Vasa}}, \bibinfo {author} {\bibfnamefont {R.}~\bibnamefont {Pomraenke}},
  \bibinfo {author} {\bibfnamefont {G.}~\bibnamefont {Cirmi}}, \bibinfo
  {author} {\bibfnamefont {E.}~\bibnamefont {De~Re}}, \bibinfo {author}
  {\bibfnamefont {W.}~\bibnamefont {Wang}}, \bibinfo {author} {\bibfnamefont
  {S.}~\bibnamefont {Schwieger}}, \bibinfo {author} {\bibfnamefont
  {D.}~\bibnamefont {Leipold}}, \bibinfo {author} {\bibfnamefont
  {E.}~\bibnamefont {Runge}}, \bibinfo {author} {\bibfnamefont
  {G.}~\bibnamefont {Cerullo}}, \ and\ \bibinfo {author} {\bibfnamefont
  {C.}~\bibnamefont {Lienau}},\ }\href {\doibase 10.1021/nn101973p} {\bibfield
  {journal} {\bibinfo  {journal} {ACS Nano}\ }\textbf {\bibinfo {volume} {4}},\
  \bibinfo {pages} {7559} (\bibinfo {year} {2010})}\BibitemShut {NoStop}%
\bibitem [{\citenamefont {Salomon}\ \emph {et~al.}(2012)\citenamefont
  {Salomon}, \citenamefont {Gordon}, \citenamefont {Prior}, \citenamefont
  {Seideman},\ and\ \citenamefont {Sukharev}}]{Salomon2012}%
  \BibitemOpen
  \bibfield  {author} {\bibinfo {author} {\bibfnamefont {A.}~\bibnamefont
  {Salomon}}, \bibinfo {author} {\bibfnamefont {R.~J.}\ \bibnamefont {Gordon}},
  \bibinfo {author} {\bibfnamefont {Y.}~\bibnamefont {Prior}}, \bibinfo
  {author} {\bibfnamefont {T.}~\bibnamefont {Seideman}}, \ and\ \bibinfo
  {author} {\bibfnamefont {M.}~\bibnamefont {Sukharev}},\ }\href {\doibase
  10.1103/PhysRevLett.109.073002} {\bibfield  {journal} {\bibinfo  {journal}
  {Phys. Rev. Lett.}\ }\textbf {\bibinfo {volume} {109}},\ \bibinfo {pages}
  {073002} (\bibinfo {year} {2012})}\BibitemShut {NoStop}%
\bibitem [{\citenamefont {DeLacy}\ \emph {et~al.}(2015)\citenamefont {DeLacy},
  \citenamefont {Miller}, \citenamefont {Hsu}, \citenamefont {Zander},
  \citenamefont {Lacey}, \citenamefont {Yagloski}, \citenamefont {Fountain},
  \citenamefont {Valdes}, \citenamefont {Anquillare}, \citenamefont {Solja{\v
  c}i{\'c}}, \citenamefont {Johnson},\ and\ \citenamefont
  {Joannopoulos}}]{DeLacy:2015ab}%
  \BibitemOpen
  \bibfield  {author} {\bibinfo {author} {\bibfnamefont {B.~G.}\ \bibnamefont
  {DeLacy}}, \bibinfo {author} {\bibfnamefont {O.~D.}\ \bibnamefont {Miller}},
  \bibinfo {author} {\bibfnamefont {C.~W.}\ \bibnamefont {Hsu}}, \bibinfo
  {author} {\bibfnamefont {Z.}~\bibnamefont {Zander}}, \bibinfo {author}
  {\bibfnamefont {S.}~\bibnamefont {Lacey}}, \bibinfo {author} {\bibfnamefont
  {R.}~\bibnamefont {Yagloski}}, \bibinfo {author} {\bibfnamefont {A.~W.}\
  \bibnamefont {Fountain}}, \bibinfo {author} {\bibfnamefont {E.}~\bibnamefont
  {Valdes}}, \bibinfo {author} {\bibfnamefont {E.}~\bibnamefont {Anquillare}},
  \bibinfo {author} {\bibfnamefont {M.}~\bibnamefont {Solja{\v c}i{\'c}}},
  \bibinfo {author} {\bibfnamefont {S.~G.}\ \bibnamefont {Johnson}}, \ and\
  \bibinfo {author} {\bibfnamefont {J.~D.}\ \bibnamefont {Joannopoulos}},\
  }\href {\doibase 10.1021/acs.nanolett.5b00157} {\bibfield  {journal}
  {\bibinfo  {journal} {Nano Lett. Article ASAP}\ } (\bibinfo {year} {2015}),\
  10.1021/acs.nanolett.5b00157}\BibitemShut {NoStop}%
\bibitem [{\citenamefont {Fedutik}\ \emph {et~al.}(2007)\citenamefont
  {Fedutik}, \citenamefont {Temnov}, \citenamefont {Sch\"ops}, \citenamefont
  {Woggon},\ and\ \citenamefont {Artemyev}}]{PhysRevLett.99.136802}%
  \BibitemOpen
  \bibfield  {author} {\bibinfo {author} {\bibfnamefont {Y.}~\bibnamefont
  {Fedutik}}, \bibinfo {author} {\bibfnamefont {V.~V.}\ \bibnamefont {Temnov}},
  \bibinfo {author} {\bibfnamefont {O.}~\bibnamefont {Sch\"ops}}, \bibinfo
  {author} {\bibfnamefont {U.}~\bibnamefont {Woggon}}, \ and\ \bibinfo {author}
  {\bibfnamefont {M.~V.}\ \bibnamefont {Artemyev}},\ }\href {\doibase
  10.1103/PhysRevLett.99.136802} {\bibfield  {journal} {\bibinfo  {journal}
  {Phys. Rev. Lett.}\ }\textbf {\bibinfo {volume} {99}},\ \bibinfo {pages}
  {136802} (\bibinfo {year} {2007})}\BibitemShut {NoStop}%
\bibitem [{\citenamefont {Schlather}\ \emph
  {et~al.}(2013{\natexlab{a}})\citenamefont {Schlather}, \citenamefont {Large},
  \citenamefont {Urban}, \citenamefont {Nordlander},\ and\ \citenamefont
  {Halas}}]{doi:10.1021/nl4014887}%
  \BibitemOpen
  \bibfield  {author} {\bibinfo {author} {\bibfnamefont {A.~E.}\ \bibnamefont
  {Schlather}}, \bibinfo {author} {\bibfnamefont {N.}~\bibnamefont {Large}},
  \bibinfo {author} {\bibfnamefont {A.~S.}\ \bibnamefont {Urban}}, \bibinfo
  {author} {\bibfnamefont {P.}~\bibnamefont {Nordlander}}, \ and\ \bibinfo
  {author} {\bibfnamefont {N.~J.}\ \bibnamefont {Halas}},\ }\href {\doibase
  10.1021/nl4014887} {\bibfield  {journal} {\bibinfo  {journal} {Nano Lett.}\
  }\textbf {\bibinfo {volume} {13}},\ \bibinfo {pages} {3281} (\bibinfo {year}
  {2013}{\natexlab{a}})}\BibitemShut {NoStop}%
\bibitem [{\citenamefont {Antosiewicz}\ \emph {et~al.}(2014)\citenamefont
  {Antosiewicz}, \citenamefont {Apell},\ and\ \citenamefont
  {Shegai}}]{doi:10.1021/ph500032d}%
  \BibitemOpen
  \bibfield  {author} {\bibinfo {author} {\bibfnamefont {T.~J.}\ \bibnamefont
  {Antosiewicz}}, \bibinfo {author} {\bibfnamefont {S.~P.}\ \bibnamefont
  {Apell}}, \ and\ \bibinfo {author} {\bibfnamefont {T.}~\bibnamefont
  {Shegai}},\ }\href {\doibase 10.1021/ph500032d} {\bibfield  {journal}
  {\bibinfo  {journal} {ACS Photon.}\ }\textbf {\bibinfo {volume} {1}},\
  \bibinfo {pages} {454} (\bibinfo {year} {2014})}\BibitemShut {NoStop}%
\bibitem [{\citenamefont {Sukharev}(2014)}]{jcp_chirps14}%
  \BibitemOpen
  \bibfield  {author} {\bibinfo {author} {\bibfnamefont {M.}~\bibnamefont
  {Sukharev}},\ }\href {\doibase http://dx.doi.org/10.1063/1.4893967}
  {\bibfield  {journal} {\bibinfo  {journal} {J. Chem. Phys.}\ }\textbf
  {\bibinfo {volume} {141}},\ \bibinfo {eid} {084712} (\bibinfo {year}
  {2014})}\BibitemShut {NoStop}%
\bibitem [{\citenamefont {Sukharev}\ \emph {et~al.}(2015)\citenamefont
  {Sukharev}, \citenamefont {Day},\ and\ \citenamefont
  {Pachter}}]{sukharev2015linear}%
  \BibitemOpen
  \bibfield  {author} {\bibinfo {author} {\bibfnamefont {M.}~\bibnamefont
  {Sukharev}}, \bibinfo {author} {\bibfnamefont {P.~N.}\ \bibnamefont {Day}}, \
  and\ \bibinfo {author} {\bibfnamefont {R.}~\bibnamefont {Pachter}},\ }\href
  {\doibase http://arxiv.org/abs/1503.07115} {\bibfield  {journal} {\bibinfo
  {journal} {arXiv preprint arXiv:1503.07115}\ } (\bibinfo {year} {2015}),\
  http://arxiv.org/abs/1503.07115}\BibitemShut {NoStop}%
\bibitem [{\citenamefont {Fernandez-Cuesta}\ \emph {et~al.}(2009)\citenamefont
  {Fernandez-Cuesta}, \citenamefont {Nielsen}, \citenamefont {Boltasseva},
  \citenamefont {Borrise}, \citenamefont {Perez-Murano},\ and\ \citenamefont
  {Kristensen}}]{Cuesta2009}%
  \BibitemOpen
  \bibfield  {author} {\bibinfo {author} {\bibfnamefont {I.}~\bibnamefont
  {Fernandez-Cuesta}}, \bibinfo {author} {\bibfnamefont {R.~B.}\ \bibnamefont
  {Nielsen}}, \bibinfo {author} {\bibfnamefont {A.}~\bibnamefont {Boltasseva}},
  \bibinfo {author} {\bibfnamefont {X.}~\bibnamefont {Borrise}}, \bibinfo
  {author} {\bibfnamefont {F.}~\bibnamefont {Perez-Murano}}, \ and\ \bibinfo
  {author} {\bibfnamefont {A.}~\bibnamefont {Kristensen}},\ }\href {\doibase
  http://dx.doi.org/10.1063/1.3262945} {\bibfield  {journal} {\bibinfo
  {journal} {Appl Phys Lett}\ }\textbf {\bibinfo {volume} {95}},\ \bibinfo
  {eid} {203102} (\bibinfo {year} {2009})}\BibitemShut {NoStop}%
\bibitem [{\citenamefont {Gramotnev}\ and\ \citenamefont
  {Pile}(2004)}]{Gramotnev2004}%
  \BibitemOpen
  \bibfield  {author} {\bibinfo {author} {\bibfnamefont {D.~K.}\ \bibnamefont
  {Gramotnev}}\ and\ \bibinfo {author} {\bibfnamefont {D.~F.~P.}\ \bibnamefont
  {Pile}},\ }\href {\doibase http://dx.doi.org/10.1063/1.1839283} {\bibfield
  {journal} {\bibinfo  {journal} {Appl Phys Lett}\ }\textbf {\bibinfo {volume}
  {85}},\ \bibinfo {pages} {6323} (\bibinfo {year} {2004})}\BibitemShut
  {NoStop}%
\bibitem [{\citenamefont {Bozhevolnyi}\ and\ \citenamefont
  {Nerkararyan}(2010)}]{Bozhevolnyi2010}%
  \BibitemOpen
  \bibfield  {author} {\bibinfo {author} {\bibfnamefont {S.~I.}\ \bibnamefont
  {Bozhevolnyi}}\ and\ \bibinfo {author} {\bibfnamefont {K.~V.}\ \bibnamefont
  {Nerkararyan}},\ }\href {\doibase 10.1364/OL.35.000541} {\bibfield  {journal}
  {\bibinfo  {journal} {Opt. Lett.}\ }\textbf {\bibinfo {volume} {35}},\
  \bibinfo {pages} {541} (\bibinfo {year} {2010})}\BibitemShut {NoStop}%
\bibitem [{\citenamefont {Bhushan}(2012)}]{EncNanotech}%
  \BibitemOpen
  \bibinfo {editor} {\bibfnamefont {B.}~\bibnamefont {Bhushan}},\ ed.,\
  \href@noop {} {\emph {\bibinfo {title} {Encyclopedia of Nanotechnology}}}\
  (\bibinfo  {publisher} {Springer},\ \bibinfo {year} {2012})\BibitemShut
  {NoStop}%
\bibitem [{\citenamefont {S\o{}ndergaard}\ and\ \citenamefont
  {Bozhevolnyi}(2009)}]{Sondergaard2009}%
  \BibitemOpen
  \bibfield  {author} {\bibinfo {author} {\bibfnamefont {T.}~\bibnamefont
  {S\o{}ndergaard}}\ and\ \bibinfo {author} {\bibfnamefont {S.~I.}\
  \bibnamefont {Bozhevolnyi}},\ }\href {\doibase 10.1103/PhysRevB.80.195407}
  {\bibfield  {journal} {\bibinfo  {journal} {Phys. Rev. B}\ }\textbf {\bibinfo
  {volume} {80}},\ \bibinfo {pages} {195407} (\bibinfo {year}
  {2009})}\BibitemShut {NoStop}%
\bibitem [{\citenamefont {Sukharev}\ \emph {et~al.}(2013)\citenamefont
  {Sukharev}, \citenamefont {Seideman}, \citenamefont {Gordon}, \citenamefont
  {Salomon},\ and\ \citenamefont {Prior}}]{Sukharev2013}%
  \BibitemOpen
  \bibfield  {author} {\bibinfo {author} {\bibfnamefont {M.}~\bibnamefont
  {Sukharev}}, \bibinfo {author} {\bibfnamefont {T.}~\bibnamefont {Seideman}},
  \bibinfo {author} {\bibfnamefont {R.~J.}\ \bibnamefont {Gordon}}, \bibinfo
  {author} {\bibfnamefont {A.}~\bibnamefont {Salomon}}, \ and\ \bibinfo
  {author} {\bibfnamefont {Y.}~\bibnamefont {Prior}},\ }\bibfield  {booktitle}
  {\emph {\bibinfo {booktitle} {ACS Nano}},\ }\href {\doibase
  10.1021/nn4054528} {\bibfield  {journal} {\bibinfo  {journal} {ACS Nano}\
  }\textbf {\bibinfo {volume} {8}},\ \bibinfo {pages} {807} (\bibinfo {year}
  {2013})}\BibitemShut {NoStop}%
\bibitem [{\citenamefont {Taflove}\ and\ \citenamefont
  {Hagness}(2005)}]{Taflove}%
  \BibitemOpen
  \bibfield  {author} {\bibinfo {author} {\bibfnamefont {A.}~\bibnamefont
  {Taflove}}\ and\ \bibinfo {author} {\bibfnamefont {S.}~\bibnamefont
  {Hagness}},\ }\href@noop {} {\emph {\bibinfo {title} {Computational
  Electrodynamics: The Finite-Difference Time-Domain Method}}}\ (\bibinfo
  {publisher} {Artech House},\ \bibinfo {year} {2005})\BibitemShut {NoStop}%
\bibitem [{\citenamefont {Kreibig}\ and\ \citenamefont {Vollmer}(1995)}]{Uwe}%
  \BibitemOpen
  \bibfield  {author} {\bibinfo {author} {\bibfnamefont {U.}~\bibnamefont
  {Kreibig}}\ and\ \bibinfo {author} {\bibfnamefont {M.}~\bibnamefont
  {Vollmer}},\ }\href@noop {} {\emph {\bibinfo {title} {Optical Properties of
  Metal Clusters}}}\ (\bibinfo  {publisher} {Springer},\ \bibinfo {year}
  {1995})\BibitemShut {NoStop}%
\bibitem [{\citenamefont {Gray}\ and\ \citenamefont {Kupka}(2003)}]{Gray2003}%
  \BibitemOpen
  \bibfield  {author} {\bibinfo {author} {\bibfnamefont {S.~K.}\ \bibnamefont
  {Gray}}\ and\ \bibinfo {author} {\bibfnamefont {T.}~\bibnamefont {Kupka}},\
  }\href {\doibase 10.1103/PhysRevB.68.045415} {\bibfield  {journal} {\bibinfo
  {journal} {Phys. Rev. B}\ }\textbf {\bibinfo {volume} {68}},\ \bibinfo
  {pages} {045415} (\bibinfo {year} {2003})}\BibitemShut {NoStop}%
\bibitem [{\citenamefont {Sukharev}\ and\ \citenamefont
  {Nitzan}(2011)}]{Sukharev2011}%
  \BibitemOpen
  \bibfield  {author} {\bibinfo {author} {\bibfnamefont {M.}~\bibnamefont
  {Sukharev}}\ and\ \bibinfo {author} {\bibfnamefont {A.}~\bibnamefont
  {Nitzan}},\ }\href {\doibase 10.1103/PhysRevA.84.043802} {\bibfield
  {journal} {\bibinfo  {journal} {Phys. Rev. A}\ }\textbf {\bibinfo {volume}
  {84}},\ \bibinfo {pages} {043802} (\bibinfo {year} {2011})}\BibitemShut
  {NoStop}%
\bibitem [{\citenamefont {Schlather}\ \emph
  {et~al.}(2013{\natexlab{b}})\citenamefont {Schlather}, \citenamefont {Large},
  \citenamefont {Urban}, \citenamefont {Nordlander},\ and\ \citenamefont
  {Halas}}]{Schlather2013}%
  \BibitemOpen
  \bibfield  {author} {\bibinfo {author} {\bibfnamefont {A.~E.}\ \bibnamefont
  {Schlather}}, \bibinfo {author} {\bibfnamefont {N.}~\bibnamefont {Large}},
  \bibinfo {author} {\bibfnamefont {A.~S.}\ \bibnamefont {Urban}}, \bibinfo
  {author} {\bibfnamefont {P.}~\bibnamefont {Nordlander}}, \ and\ \bibinfo
  {author} {\bibfnamefont {N.~J.}\ \bibnamefont {Halas}},\ }\href {\doibase
  10.1021/nl4014887} {\bibfield  {journal} {\bibinfo  {journal} {Nano Lett}\
  }\textbf {\bibinfo {volume} {13}},\ \bibinfo {pages} {3281} (\bibinfo {year}
  {2013}{\natexlab{b}})},\ \bibinfo {note} {pMID: 23746061}\BibitemShut
  {NoStop}%
\bibitem [{\citenamefont {Hutchison}\ \emph {et~al.}(2011)\citenamefont
  {Hutchison}, \citenamefont {O'Carroll}, \citenamefont {Schwartz},
  \citenamefont {Genet},\ and\ \citenamefont {Ebbesen}}]{Hutchison2011}%
  \BibitemOpen
  \bibfield  {author} {\bibinfo {author} {\bibfnamefont {J.~A.}\ \bibnamefont
  {Hutchison}}, \bibinfo {author} {\bibfnamefont {D.~M.}\ \bibnamefont
  {O'Carroll}}, \bibinfo {author} {\bibfnamefont {T.}~\bibnamefont {Schwartz}},
  \bibinfo {author} {\bibfnamefont {C.}~\bibnamefont {Genet}}, \ and\ \bibinfo
  {author} {\bibfnamefont {T.~W.}\ \bibnamefont {Ebbesen}},\ }\href {\doibase
  10.1002/anie.201006019} {\bibfield  {journal} {\bibinfo  {journal} {Angew
  Chem Int Edit}\ }\textbf {\bibinfo {volume} {50}},\ \bibinfo {pages} {2085}
  (\bibinfo {year} {2011})}\BibitemShut {NoStop}%
\bibitem [{\citenamefont {Sugawara}\ \emph {et~al.}(2006)\citenamefont
  {Sugawara}, \citenamefont {Kelf}, \citenamefont {Baumberg}, \citenamefont
  {Abdelsalam},\ and\ \citenamefont {Bartlett}}]{Sugawara2006}%
  \BibitemOpen
  \bibfield  {author} {\bibinfo {author} {\bibfnamefont {Y.}~\bibnamefont
  {Sugawara}}, \bibinfo {author} {\bibfnamefont {T.~A.}\ \bibnamefont {Kelf}},
  \bibinfo {author} {\bibfnamefont {J.~J.}\ \bibnamefont {Baumberg}}, \bibinfo
  {author} {\bibfnamefont {M.~E.}\ \bibnamefont {Abdelsalam}}, \ and\ \bibinfo
  {author} {\bibfnamefont {P.~N.}\ \bibnamefont {Bartlett}},\ }\href {\doibase
  10.1103/PhysRevLett.97.266808} {\bibfield  {journal} {\bibinfo  {journal}
  {Phys. Rev. Lett.}\ }\textbf {\bibinfo {volume} {97}},\ \bibinfo {pages}
  {266808} (\bibinfo {year} {2006})}\BibitemShut {NoStop}%
\bibitem [{\citenamefont {Agranovich}\ and\ \citenamefont
  {La~Rocca}(2005)}]{Agranovich2005}%
  \BibitemOpen
  \bibfield  {author} {\bibinfo {author} {\bibfnamefont {V.}~\bibnamefont
  {Agranovich}}\ and\ \bibinfo {author} {\bibfnamefont {G.}~\bibnamefont
  {La~Rocca}},\ }\href@noop {} {\bibfield  {journal} {\bibinfo  {journal}
  {Solid State Commun}\ }\textbf {\bibinfo {volume} {135}},\ \bibinfo {pages}
  {544} (\bibinfo {year} {2005})}\BibitemShut {NoStop}%
\bibitem [{\citenamefont {Allen}\ and\ \citenamefont {Eberly}(1975)}]{Eberly}%
  \BibitemOpen
  \bibfield  {author} {\bibinfo {author} {\bibfnamefont {L.}~\bibnamefont
  {Allen}}\ and\ \bibinfo {author} {\bibfnamefont {J.~H.}\ \bibnamefont
  {Eberly}},\ }\href@noop {} {\emph {\bibinfo {title} {Optical Resonance and
  Two-Level Atoms}}}\ (\bibinfo  {publisher} {John Wiley \& Sons, Inc.},\
  \bibinfo {year} {1975})\BibitemShut {NoStop}%
\bibitem [{\citenamefont {Wang}\ \emph {et~al.}(2000)\citenamefont {Wang},
  \citenamefont {Kuzmich},\ and\ \citenamefont {Dogariu}}]{Wang2000}%
  \BibitemOpen
  \bibfield  {author} {\bibinfo {author} {\bibfnamefont {L.~J.}\ \bibnamefont
  {Wang}}, \bibinfo {author} {\bibfnamefont {A.}~\bibnamefont {Kuzmich}}, \
  and\ \bibinfo {author} {\bibfnamefont {A.}~\bibnamefont {Dogariu}},\ }\href
  {http://dx.doi.org/10.1038/35018520} {\bibfield  {journal} {\bibinfo
  {journal} {Nature}\ }\textbf {\bibinfo {volume} {406}},\ \bibinfo {pages}
  {277} (\bibinfo {year} {2000})}\BibitemShut {NoStop}%
\bibitem [{\citenamefont {Siddiqui}\ and\ \citenamefont
  {Mojahedi}(2003)}]{Siddiqui2003}%
  \BibitemOpen
  \bibfield  {author} {\bibinfo {author} {\bibfnamefont {O.~F.}\ \bibnamefont
  {Siddiqui}}\ and\ \bibinfo {author} {\bibfnamefont {M.}~\bibnamefont
  {Mojahedi}},\ }\href@noop {} {\bibfield  {journal} {\bibinfo  {journal} {IEEE
  T Antenn Propag}\ }\textbf {\bibinfo {volume} {51}},\ \bibinfo {pages} {2619}
  (\bibinfo {year} {2003})}\BibitemShut {NoStop}%
\bibitem [{\citenamefont {Woodley}\ and\ \citenamefont
  {Mojahedi}(2003)}]{Mojahedi2003}%
  \BibitemOpen
  \bibfield  {author} {\bibinfo {author} {\bibfnamefont {J.}~\bibnamefont
  {Woodley}}\ and\ \bibinfo {author} {\bibfnamefont {M.}~\bibnamefont
  {Mojahedi}},\ }in\ \href@noop {} {\emph {\bibinfo {booktitle} {IEEE Antennas
  Prop}}}\ (\bibinfo {year} {2003})\ pp.\ \bibinfo {pages} {643--646 vol.
  4}\BibitemShut {NoStop}%
\bibitem [{\citenamefont {Brixner}\ and\ \citenamefont
  {Gerber}(2003)}]{Brixner:2003aa}%
  \BibitemOpen
  \bibfield  {author} {\bibinfo {author} {\bibfnamefont {T.}~\bibnamefont
  {Brixner}}\ and\ \bibinfo {author} {\bibfnamefont {G.}~\bibnamefont
  {Gerber}},\ }\href {\doibase 10.1002/cphc.200200581} {\bibfield  {journal}
  {\bibinfo  {journal} {ChemPhysChem}\ }\textbf {\bibinfo {volume} {4}},\
  \bibinfo {pages} {418} (\bibinfo {year} {2003})}\BibitemShut {NoStop}%
\end{thebibliography}
\end{document}